\documentclass[a4paper,11pt]{article}  
\usepackage[T1]{fontenc}
\usepackage{makeidx}

\usepackage{amssymb}

\usepackage[dvips]{graphicx}

\newcounter{prgline} 

\newcommand{\citelist}[1]{\raisebox{.2ex}{[}#1\raisebox{.2ex}{]}}
\newcommand{\scite}[1]{\citeauthor{#1}, \citeyear{#1}}

\newcommand{\multiciteiii}[3]%
  {\citelist{\scite{#1}, \citeyear{#2}, \citeyear{#3}}}

\renewcommand{\phi}{\varphi}
\renewcommand{\epsilon}{\varepsilon}

\newcommand{\text}[1]{\mbox{\rm \,#1\,}}        

\newcommand{\pushlist}[1]{\setcounter{#1}{\value{enumi}} \end{enumerate}}
\newcommand{\poplist}[1]{\begin{enumerate} \setcounter{enumi}{\value{#1}}}

\renewcommand{\emptyset}{\varnothing}  
\newtheorem{defnx}{Definition}

\newtheorem{theoremx}[defnx]{Theorem}

\newtheorem{lemmax}[defnx]{Lemma}

\newcommand{\QED}{\nopagebreak[4]{\makebox[1mm]{}\hfill$\square$}}
\newenvironment{theorem}{\begin{theoremx}\em}{\end{theoremx}}
\newenvironment{lemma}{\begin{lemmax}\em}{\end{lemmax}}
\newenvironment{proof}{\noindent {\bf \noindent Proof: }}{\QED \\}

\newlength{\prgindent}
\newenvironment{program}{
  \begin{list}%
   {\arabic{prgline}}%
   {
   \usecounter{prgline}
   \setlength{\prgindent}{0em}
   \setlength{\parsep}{0em}
   \setlength{\itemsep}{0em}
   \setlength{\labelwidth}{1em}
   \setlength{\labelsep}{1em}
   \setlength{\leftmargin}{\labelwidth}
   \addtolength{\leftmargin}{\labelsep}
   \setlength{\topsep}{0em}
   \setlength{\parskip}{0em} } }%
 {\end{list}}

\newif\ifprgendtext
\prgendtexttrue

{\addtolength{\prgindent}{-\labelsep}}
\newcommand{\prgbeginblock}{\addtolength{\prgindent}{\labelsep}}
\newcommand{\prgendblock}{\addtolength{\prgindent}{-\labelsep}}
\newcommand{\prgcndendblock}[1]{\addtolength{\prgindent}{-\labelsep}
 \ifprgendtext \prglin\prgres{#1}\fi}

\newcommand{\prglin}{\rm \item\hspace{\prgindent}}

\newcommand{\prgres}[1]{{\bf #1}}

\newcommand{\prgif}{\prglin\prgres{if\ }}
\newcommand{\prgthen}{\prgres{\ then\ }\prgbeginblock}
\newcommand{\prgelse}{\prgendblock\prglin\prgres{else}\prgbeginblock}
\newcommand{\prgelsif}{\prgendblock\prglin\prgres{elsif\ }}
\newcommand{\prgendif}{\prgcndendblock{end if}}

\newcommand{\prgfor}{\prglin\prgres{for\ }}
\newcommand{\prgdo}{\prgres{\ do}\prgbeginblock}

\newcommand{\prgendfor}{\prgcndendblock{end for}}

\newcommand{\prgand}{\prgres{\ and\ }}
\newcommand{\prgor}{\prgres{\ or \ }}

\newcommand{\prgreturn}{\prgres{return\ }}

\prgendtextfalse

\newcommand{\ie}{{i.e.}}                
\newcommand{\eg}{{e.g.}}

\newcommand{\unprint}[1]{}

\newcommand{\xsat}{{\sc xsat }}

\newcommand{\xsatnosp}{{\sc xsat}}

\newcommand{\hamcsp}{{\sc max hamming csp }}

\newcommand{\hamxsat}{{\sc max hamming xsat }}
\newcommand{\hamxsatnosp}{{\sc max hamming xsat}}

\newcommand{\hamsat}{{\sc max hamming sat }}
\newcommand{\hamsatnosp}{{\sc max hamming sat}}

\title{Algorithms for Max Hamming Exact Satisfiability}

\author{Vilhelm Dahll{\"o}f\thanks{The research is supported by CUGS -- National
 Graduate School in Computer Science, Sweden.}\\
Dept. of Computer and 
Information Science\\ 
Link{\"o}ping University\\
SE-581 83 Link{\"o}ping,  Sweden\\ {\tt vilda@ida.liu.se}
}

\begin{document}

\date{}

\maketitle

\begin{abstract}
We here study \hamxsatnosp, \ie, the problem of finding two \xsat models at maximum Hamming distance. By using a recent \xsat solver as an auxiliary function, an $O(2^n)$ time algorithm can be constructed, where $n$ is the number of variables. This upper time bound can be further improved to $O(1.8348^n)$ by introducing a new kind of branching, more directly suited for finding models at maximum Hamming distance. The techniques presented here are likely to be of practical use as well as of theoretical value, proving that there are non-trivial algorithms for maximum Hamming distance problems.
\end{abstract}

\bibliographystyle{abbrv}

\section{Introduction}
Most previous algorithms for optimization problems have contented themselves with producing {\em one} best or good-enough solution. However, often there is an actual need for {\em several solutions} that are at a maximum (or at least great) Hamming distance. For instance, when scheduling a group of people one typically wants to present substantially different alternatives to choose between. Somewhat surprisingly, the \hamcsp problem has only recently become an area of academic research. The first paper (to the best of our knowledge) by Rossi {\em et al.}~\cite{Rossi-02} came in 2002. In their paper they present some results on the hardness of approximating the problem for CSPs on Boolean domains. Angelsmark and Thapper \cite{Angelsmark-04} have presented exact and randomized algorithms for the general finite domain problem as well as dedicated algorithms for \hamsatnosp. Hebrard {\em et al.} \cite{Hebrard-05} consider a broader range of problems, including finding solutions that are similar. They also test some heuristic methods. The so far best exact algorithm for \hamsat by Angelsmark and Thapper \cite{Angelsmark-04} runs in $O(4^n)$ time (where $n$ is the number of variables) and polynomial space.


In this paper we will consider \hamxsatnosp. The \xsat problem asks for an assignment to the variables such that exactly one literal be true in each clause. \xsat is NP-complete as shown by Schaefer \cite{Schaefer-78}. The problem is well studied, and many exact algorithms have been presented, \eg~\cite{Peleg-TCS-02,Hirsch:Kulikov:2002,Porschen:2002,Dahllof-TCS-04,Byskov-TCS-04}. The so far best algorithm by Byskov {\em et al.}~\cite{Byskov-TCS-04} have a running time in $O(1.1749^n)$ and uses polynomial space.  \xsat can be used to model for instance the graph colourability problem (since every vertex must have exactly one colour, for an example see \cite{Liu:CP:03}). Furthermore, there is a close connection between \xsat and more general cardinality constraints (see \cite{Dahllof-04}).  \hamxsat is not efficiently approximable (see \cite{Rossi-02}) and so exact algorithms are of real-world interest. 

We will present two polynomial space algorithms $P$ and $Q$. Previous algorithms for maximum Hamming problems have relied on an external solver for the base problem. $P$ is also such an algorithm, however, there is a novelty: by using a polynomial time test, many unnecessary calls to the solver can be avoided. Thereby the running time is improved substantially. $Q$ represents something totally new in this area, because it works directly on the inherent structure of the \hamxsat problem. More precisely, a new kind of DPLL branching is introduced. Apart from the immediate interest of the \hamxsat problem itself, we hope that the ideas presented here will also be applicable for other problems such as {\sc max hamming scheduling}, {\sc max hamming clique} and the like.

For the sake of conciseness, we phrase the algorithms in such a way that they answer the question ``what is the maximum Hamming distance between any two models?''. However, it is trivial to see how they can be modified to actually produce two such models.

In what follows we first give some preliminaries and then in Section 3 we present $P$ and $Q$. In Section 4 some conclusions and possible future research directions are given.

\section{Preliminaries}
A {\em propositional variable} (or {\em variable} for short) has
either the value $true$ or $false$. A {\em literal} is a variable $p$
or its negation $\bar p$. We say that the literals $p$ and $\bar p$ are {\em derived} from the variable $p$. When {\em flipping} $p$ ($\bar p$) one gets $\bar p$ ($p$). The literal $p$ is $true$ iff it is derived from the variable $p$ which has the value $true$ and $\bar p$ is $true$
iff it is derived from the variable $p$ which has the value $false$.
 A {\em  clause} is a number of literals connected by logical or
($\lor$).  The {\em length} of a clause $x$, denoted $|x|$, is the
number of literals in it. We will sometimes need a sub-clause notation in this
way: $(a \lor b \lor C)$, such that $C= c_0 \lor \ldots \lor c_n$ is a
disjunction of one or more literals.  In the following, literals will be
indicated by lower-case letters and sub-clauses by upper-case letters. 
A {\em formula} is a set of clauses. For a formula $F$, $Var(F)$ denotes the set of variables appearing in a clause of $F$.
  The {\em degree of $c$}, denoted $\delta(c)$, is the number of appearances of the variable $c$, that is, the number of clauses that contain either
$c$ or $\bar c$. If $\delta(c)=1$ we call $c$ a {\em singleton}. From a formula one gets the {\em formula graph} by letting the variables form the vertices and every pair of variables occuring together in a clause is joined by an edge. Hence, graph concepts such as ``connected components'' can be used for formulae.

An {\em x-model} is an assignment to the variables of a formula $F$ such that there is exactly one true literal in every clause. The problem of determining whether $F$ allows an x-model is called \xsatnosp. A literal that exactly satisfies a clause is called a {\em satisfactor}. 

We now reach two central definitions: The {\em Hamming distance} between two assignments is the number of assignments to the individual variables that disagree. \hamxsat is the problem of determining for a formula $F$ the maximum Hamming distance between any two x-models of $F$. 

{\em Substitution} of $a$ by $\delta$ in the formula $F$ is denoted $F(a/\delta)$; the notation $F(a/\delta;b/\gamma)$ indicates repeated substitution: $F(a/\delta)(b/\gamma)$ (first $a$ is replaced and then $b$). $F(B/false)$ means that every literal of $B$ is replaced by $false$. 

We will deal with variants of the \xsat problem, and in order not to clutter the algorithms with trivialities, we shall assume that the substitution  performs a little more than just a syntactical replacement, namely propagation in the following sense: Given a formula $F$, assume that there are three clauses $x=(a \lor b \lor c),y=(b \lor f \lor g \lor h)$ and $z=(\bar c \lor d \lor e)$ in a formula. When substituting true for $a$ ($F(a/true)$), $b$ and $c$ must both be replaced by false (because in the context of \xsat exactly one literal must be true in each clause). This means that $y$ will become $(false \lor f \lor g \lor h)$ which can be simplified to $(f \lor g \lor h)$ and that $z$ will become $(true \lor d \lor e)$ which implies that $d$ and $e$ are false, and so on. Other trivial simplifications are also made. For instance, the occurrence of both $a$ and $\bar a$ in a clause is replaced by true. This process goes on until no more simplifications can be done. If the substitution discovers that the formula is x-unsatisfiable (for instance if there is a clause $(true \lor  a \lor \bar a)$) the unsatisfiable formula $\{()\}$ is returned.

When analyzing the running time of the algorithms, we will encounter recurrences of the form $T(n) \leq \sum_{i=1}^k T(n-r_i)+{\rm poly}(n).$
They satisfy $T(n) \in O(\tau(r_1,\ldots,r_k)^n)$
where $\tau(r_1,\ldots,r_k)$ is the largest, real-valued
root of the function 

\begin{equation} \label{rot}
f(x)=1-\sum_{i=1}^k x^{-r_i}
\end{equation} 

\noindent see \cite{Kullman-99}.
Since this bound does not depend on the polynomial factor
${\rm poly}(n)$, we  ignore all 
polynomial-time calculations. Let  $R=\sum_{i=1}^kr_i$ and then note that due to the nature of the function $f(x)=1-\sum_{i=1}^k x^{-r_i}$, the smallest possible real-valued root (and hence the best running time) will appear when each $r_i$ is as close to $R/k$ as possible, \ie, when the decrease of size of the instance is balanced through the branches. Say for instance that $R=4, k=2$. Then $\tau(1,3) = \tau(3,1) \approx 1.4656$ and $\tau(2,2) \approx 1.4142$. We will refer to this as {\em the balanced branching effect.} We will use the shorthand notation $\tau(r^k \ldots)$ for $\tau(\underbrace{r,r \ldots r}_k, \ldots)$, \eg, $\tau(5^2,3^3)$ for $\tau(5,5,3,3,3)$. 

\section{Exact Algorithms for \hamxsat}
In what follows we present the two poly-space algorithms $P$ and $Q$ for \hamxsat and prove that they run in $O(2^n)$ and $O(1.8348^n)$ time respectively. Though the running time of $P$ is slightly inferior to the running time of $Q$, there are good reasons to present both algorithms: $P$ resembles previous algorithms and gives a hint on how they can be improved, and it is easy to implement given an external \xsat solver. Furthermore, if one is content with getting two models that have at least the Hamming distance $d$, for some constant $d$, then $P$ will have a provably better upper time bound than $Q$.

As a convention, when we present a clause $(a \lor \ldots)$, it is intended to cover all dual cases as well, \ie, $(\bar a \lor \ldots)$.

\subsection{Using an External \xsat Solver}
One solution to the \hamxsat problem is this algorithm which bears resemblance to the $O(4^n)$ time \hamsat algorithm by Angelsmark and Thapper \cite{Angelsmark-04}. If the formula is x-unsatisfiable $\bot$ is returned. The answer 0 of course indicates that there is only one model.

\begin{program}
\prglin {\bf algorithm} $P(F)$
\prglin $ans:=\bot$
\prgfor $k:=0$ to $n$ \prgdo

 \prgfor every subset $X \subseteq Var(F)$ of size $k$ \prgdo
  
  \prglin Let $C$ be the set of clauses containing any literal derived from $X$
  \prglin Let $C'$ be a copy of $C$ where every literal derived from $X$ is flipped 
  \prgif  all clauses of $C$ contain either 0 or 2 literals form $X$ \prgthen
  \prgif  $F \cup C'$ is x-satifiable \prgthen $ans:=k$
  \prgendif
  \prgendif
 \prgendfor

\prgendfor

\prglin \prgreturn $ans$
\end{program}

$\textrm{ }$

\noindent Before stating the correctness of $P$ we need an auxiliary lemma.

\begin{lemma}\label{zeroTwo}
Assume that $M$ and $M'$ are x-models for $F$ and that $X$ is the subset of variables assigned different values. Then each clause of $F$ contains either zero or two literals derived from $X$.
\end{lemma}

\begin{proof}
For $M=M'$ the Lemma trivially holds.

Else, for the sake of contradiction, assume there is a clause having one literal $a$ from $X$. The clause cannot be $(a)$ because then it would be unsatisfied under one model. Therefore the clause must be $(a \lor A)$ where all members of $A$ have the same value under both models. If one literal of $A$ is true then $a$ must be false under both models (to avoid oversatisfaction), clearly a contradiction. If all literals of $A$ are false, $a$ must be false so this also is a contradiction.

Similarly, no clause can contain three or more literals from $X$. 
\end{proof}

\begin{theorem}
$P(F)$ decides \hamxsat for $F$
\end{theorem}

\begin{proof}
{\em For completeness:} Assume there are two models $M$ and $M'$ at maximum hamming distance $k$ and that the differing variables are collected in $X$. The clauses containing zero literals from $X$ remain the same under both models, the interesting case is a clause $(a \lor b \lor C)$ where $a$ and $b$ are from $X$ (by Lemma \ref{zeroTwo} this is the only possible case). Assume w.l.o.g.~that $a$ is true and $b$ is false under $M$ and the opposite holds for $M'$. Then the clause $(\bar a \lor \bar b \lor C)$ is x-satisfied under both models.

{\em For soundness:} Assume we have a model $M$ for $F \cup C'$. Then it is possible to form another model $M'$ by assigning all variables of $X$ the opposite values. 
\end{proof}

We can now start examining the running time of $P$. Let an {\em allowed subset} $S$ of variables in a formula $F$ be a subset such that each clause of $F$ contains either 0 or two members of $S$. The following lemma establishes an upper bound for the number of allowed subsets. 

\begin{lemma}\label{allowed}
For any formula $F$ the number $N$ of allowed subsets is in $O(7^{n/4}) \subseteq O(1.6266^n)$. 
\end{lemma}

\begin{proof}
Consider a arbitrary variable $a$. When calculating the number $N$ of allowed subsets $a$ can participate in, it is clear that the higher the degree of $a$, the lower the $N$. Hence, a formula consisting only of singletons has maximum $N$. 

Which clause length $l$ maximizes $N$? We see that $N=({l \choose 2}+1)^{n/l}$. Clause length 2 makes $N \in O(2^{n/2}) \subseteq O(1.4143^n)$; length 3 makes  $N \in O(4^{n/3}) \subseteq O(1.5875^n)$; length 4 makes  $N \in O(7^{n/4}) \subseteq O(1.6266^n) $; length 5 makes  $N \in O(11^{n/5}) \subseteq O(1.6154^n)$;  length 6 makes  $N \in O(16^{n/6}) \subseteq O(1.5875^n)$ and so on in a decreasing series. (The series decreases asymptotically towards 1 because the base increases only quadratically under an exponential decrease.) Thus, the maximum $N$ is in $O(7^{n/4})$.
\end{proof}

\begin{theorem}
$P(F)$ runs in polynomial space and time $O(2^n)$.
\end{theorem}

\begin{proof}
Clearly $P$ uses polynomial space. Furthermore, the running time is $O(2^n + N \cdot C^n)$, where $N$ is the constant of Lemma \ref{allowed} and $C$ is a constant such that \xsat is solvable in polynomial space and time $O(C^n)$. The currently best value for $C$ is 1.1749 (by Byskov {\em et al.}, \cite{Byskov-TCS-04}) and so the upper time bound is $O(2^n + 1.6266^n1.1749^n) \subseteq O(2^n + 1.9111^n) = O(2^n)$.
\end{proof}

\subsection{Using Branching}

We will now move on to another poly-space algorithm $Q$ with a provably better running time than $P$. It is a DPLL-style algorithm relying on the fact that under two models $M$ and $M'$, any variable $a$ has either the same or opposite value. If $a$ is true under both models, then all variables occuring in a clause $w=(a \lor \ldots)$ can be removed (because only one literal is true). If $a$ is false under both models it can be removed. If $a$ has different values then by Lemma \ref{zeroTwo}  there is exactly one more variable $a'$ in $w$ that has different values and we need to examine all possible cases of $a'$. During the branching some simplifications of the formula are made, for instance, superfluous singletons are removed. 
We need to store information about removals of variables due to simplifications and therefore the following is needed: To every variable $a$ we associate two (possibly empty) sets of variables: $sing(a)$ and $dual(a)$. We also need a marker $sat(a)$. 
 As a consequence of the simplifications, in the leaves of the recursion tree a kind of generalized models are found, that summarize several models. For now, we hide the details in the helper algorithm $Gen_H$ which we will come back to after the presentation of the main algorithm. The reason for doing so, is that we first need to see how the simplifications work.

Another technicality: like $P$, $Q$ may return $\bot$ if $F$ is unsatisfiable. Therefore we define $\bot < 0$ and $\bot+1=\bot$; furthermore, $\max_\bot(\bot, Z)$ returns $Z$, even if $Z=\bot$.
Before $Q'(F)$ is used, all sets $dual(a)$ and $sing(a)$ are assumed to be empty, and every marker $sat(a)$ assumed to be unassigned. During the execution of $Q'$, if there is a clause $(a \ldots)$ where $a$ is a singleton assigned a satisfactor, then $sat(a):=true$, in the dual case where the clause looks like $(\bar a \ldots)$, $sat(a):=false$. This allows us to find out the role of $a$ in a model.

For clarity of presentation we will first present a simplified algorithm $Q'$. Later an optimization to improve the running time will be added.


\begin{program}
\prglin {\bf algorithm} $Q'(F)$
 \prglin As long as there is a clause $(a_1 \lor a_2 \ldots)$ where $a_1$ and $a_2$ are singletons, remove $a_2$ and let $sing(a_1):=sing(a_1) \cup \{a_2\} \cup sing(a_2)$.
 \prglin  As long as there is a clause $(a \lor b)$, assume w.l.o.g.~that $b$ is a non-singleton (otherwise pick $a$) and let $F:=F(a/\bar b)$ and let $dual(b):= dual(b) \cup dual(a) \cup \{a\}$. If a singleton was created, goto the previous line.

 \prgif $F=\{()\}$ \prgthen \prgreturn $\bot$
 \prgelsif $F=\{\}$ \prgthen \prgreturn $Gen_H(F)$

 \prgelsif $F$ is not connected \prgthen assume the components are $F_1 \ldots F_k$ and return $\sum_{i=1}^k Q'(F_i)$

 \prgelse 
  \prglin Pick a longest clause $w=(a_1 \lor a_2 \ldots a_k)$ and assume w.l.o.g.~that $a_1$ is a non-singleton. Now do the following:

\prglin $ans_{true}:=Q'(F(a_1/true))$
\prglin $ans_{false}=Q'(F(a_1/false))$
\prgif  $ans_{true}=\bot$ \prgor $ans_{false}=\bot$ \prgthen \prgreturn $\max_\bot(ans_{true},ans_{false})$ \prgelse
\prgfor $i=2$ to $k$ \prgdo
\prglin Let $ans_{i}:=Q'(F(a_1/\bar a_i))$
\prgendfor
\prglin \prgreturn $\max_\bot(ans_{true},ans_{false}, (ans_2 + 1) \ldots, (ans_k + 1))$
\prgendif
\prgendif
\end{program}

$\textrm{ }$

We are now ready to take a closer look at how the result of the simplifications are handled by $Gen_H$. Note that during the execution of $Q'$, every removed variable is kept in exactly one set $sing(a)$ or $dual(a)$, for (possibly) different variables $a$. This motivates the following definition:

A {\em generalized assignment} is a partial assignment, such that every unassigned variable is contained in exactly one set $sing(a)$ or $dual(a)$ (\ie, for all the sets $sing(a_1), dual(a_1), sing(a_2) \ldots$, every intersection is empty). We say that a variable $a'$ is {\em transitively linked} to the variable $a$ if either 1) $a' \in sing(a) \cup dual(a)$ or 2) $a$ is transitively linked to a member of $sing(a) \cup dual(a)$.

We will also need the two following auxiliary algorithms. Intuitively, $Fix(a_1)$ corresponds to the maximum number of variables transitively linked to $a_1$ that can have different values under a model $M$ where $a_1$ is a satisfactor and a model $M'$ where $a_1$ is not a satisfactor.  The recursive algorithm $di(a_1)$ calculates the maximum number of variables, transitively linked to $a_1$ that can be assigned different values while $a_1$ is a non-satisfactor. In the recursive calls, it might be that the argument is a satisfactor. The variable $k$ is assumed to be initialized to $0$. 

$\textrm{ }$

\begin{program}
\prglin {\bf algorithm} $Fix(a_1)$
\prglin $fix:=0$
\prgif $sing(a) \neq \emptyset$ \prgthen
\prglin Let $\{a_1,a_2 \ldots a_m\}:=\{a_1\} \cup sing(a_1)$
\prglin $sing(a_1):=\emptyset$
\prglin $fix:=\max(Fix(a_1),Fix(a_2) \ldots Fix(a_m))$
\prgelsif $dual(a) \neq \emptyset$ \prgthen
\prglin Let $\{a_1,a_2 \ldots a_m\}:=\{a_1\} \cup dual(a_1)$
\prglin $dual(a_1):=\emptyset$
\prglin $fix:=\sum(Fix(a_1),Fix(a_2) \ldots Fix(a_m))$
\prgelse 
\prglin $fix:=1$
\prgendif
\prglin \prgreturn $fix$ 
\end{program}

$\textrm{ }$

\begin{program}
\prglin {\bf algorithm} $di(a_1)$

\prgif $sing(a_1) \neq \emptyset$ \prgand $a_1$ is a satisfactor \prgthen
\prglin $k:= k + Fix(a_1)$
\prgelse
\prgfor each member $b_i \in dual(a_1) \cup sing(a_1)$ \prgdo
\prglin assign $b_i$ a value according to $a_1$; $k:=k+di(b_i)$
\prgendfor
\prgendif

\prglin \prgreturn $k$ 
\end{program}

$\textrm{ }$

We are now ready to present $Gen_H(F)$. Although $F$ is an empty formula, it is assumed that from it, every variable assigned a value during the execution of $Q'$ can be reached. 

$\textrm{ }$

\begin{program}
\prglin {\bf algorithm} $Gen_H(F)$
\prglin $k:=0$
\prgfor every variable $a_1$ assigned a value \prgdo

\prgif $sing(a_1)=\emptyset$ and $dual(a_1) = \emptyset$ \prgthen do nothing

\prgelsif $sing(a_1) = \{a_2 \ldots a_m\}$ and $a_1$ is a satisfactor \prgthen
\prglin Pick two members $a'$ and $a''$ from $\{a_2 \ldots a_m\}$, there are ${ m \choose 2}$ choices. Try all and for each choice calculate $k_1:=Fix(a') + Fix(a'') + \sum di(a_i)$ such that $a_i \in \{a_1 \ldots a_m\} \setminus \{a' \cup a''\}$. The maximum $k_1$ found is added to $k$.

\prgelse

\prglin $k:=k+di(a_1)$

\prgendif

\prgendfor

\prglin \prgreturn $k$ 
\end{program}

$\textrm{ }$

\noindent We are now ready to state the correctness of $Q'$:

\begin{theorem}
$Q'(F)$ decides \hamxsat for $F$
\end{theorem}

\begin{proof}
We inspect the lines of $Q'$:

\noindent {\bf Lines 2--5:} Let us start by looking at Lines 2 and 3 to see that they do not alter the x-satisfiability of $F$ and that they indeed produce a generalized assignment. As for Line 2, it is clear that removing all singletons but one does not alter the x-satisfiability. It is also clear that every removed singleton will be in one and only one set $sing$. Concerning Line 3, the clause $(a \lor b)$ implies that $a$ and $b$ have opposite values, hence $F:=F(a/\bar b)$ does not alter the x-satisfiability. By the previous line, one of $a$ and $b$ is a non-singleton and so every variable removed by this line is found in exactly one set $dual$. The formula $\{()\}$ is unsatisfiable and thus $\bot$ is returned. When it comes to $Gen_H$, we need to justify the calculation of the maximum Hamming distance for a generalized assignment. Consider two models $M$ and $M'$ at maximum Hamming distance, contained in the generalized assignment at hand. Clearly, all variables that are assigned a fixed value and have empty sets $sing$ and $dual$ will have the same value under both models. Furthermore, whenever there is a situation with a satisfactor $a_1$ having a non-empty set $sing(a_1)$, one of $a_1 \ldots a_m$ will be a satisfactor under $M$ and one under $M'$. When $a_i$ is a satisfactor under $M$, $Fix(a_1)$ is the largest number of variables transitively linked to it that can get assigned one value under $M$ and another value under $M'$. Also, even though a variable is not a satisfactor itself, it may well be that one variable transitively linked to it is. As in a general assignment every variable is either assigned a fixed value or transitively linked to such a variable, the distance between $M$ and $M'$ can be found as the sum of the values calculated for the assigned variables. 

\noindent {\bf Line 6:} If $F$ is not connected every model for one component can be combined with any model for another component in order to form a model for $F$.

\noindent {\bf Lines 7--15:} Assume there are two models $M$ and $M'$ at maximum Hamming distance $k$. If $a_1$ is true under both models then the formula where all other literals of $w$ are set to false is x-satisfiable and the recursive call will return $k$ (assuming that the algorithm is correct for smaller input). Similarly for Line 10. If both Lines 9 and 10 returned an integer, we know that there are models under which $a_1$ is false and models under which $a_1$ is true. Thus $M$ and $M'$ may assign different values to $a_1$. Assume this is the case. By Lemma \ref{zeroTwo} we know that $M$ and $M'$ differ in exactly one more variable in $w$. Assume w.l.o.g.~that $a_2$ is that literal. Then we know that $a_1$ and $a_2$ have different values and that the other literals of $w$ are false.

\end{proof}

As for  the running time of $Q'$, the handling of clauses of length 4 will cause an unnecessarily bad upper time bound. The problem is that in Line 10 only one variable is removed. However, a clause of length 3 is created which can be exploited. Hence we replace Line 10 in $Q'$ by the following, thereby obtaining the algorithm $Q$.  The correctness is easily seen, because it is the same kind of branching we have already justified.

$\textrm{ }$

\begin{program}
\prgif $|W| \neq 4$ \prgthen $ans_{false}=Q(F(a_1/false))$
\prgelse 
\prglin let $W=(a_1 \lor a_2 \lor a_3 \lor a_4)$ and assume that $a_2$ is a non-singleton
\prglin $ans^1_{f}:=Q(F(a_1/false;a_2/true)$; $ans^2_{f}:=Q(F(a_1/false;a_2/false)$

\prgif $ans^1_{f}=\bot$ \prgor $ans^2_{f}=\bot$ \prgthen $ans_{false}:=\max_\bot(ans^1_{f},ans^2_{f})$

\prgelse 

\prglin $ans^3_{f}:=Q(F(a_1/false;a_2/\bar a_3)$; $ans^4_{f}:=Q(F(a_1/false;a_2/\bar a_4)$

\prglin  $ans_{false}:=\max_\bot(ans^1_{f},ans^2_{f},(ans^3_{f}+1),(ans^4_{f}+1))$

\prgendif

\prgendif
\end{program}

$\textrm{ }$

\begin{theorem}
$Q(F)$ runs in polynomial space and time $O(1.8348^n)$
\end{theorem}

\begin{proof}
Let $T(n)$ be the running time for $Q(F)$. The analysis will proceed by examining what the running time would be if $Q$ always encountered the same case. It is clear that the worst case will decide an overall upper time bound for $Q$.
We inspect the lines of $Q$: 

\noindent {\bf Line 1--5:} All these lines are polynomial time computable.

\noindent {\bf Line 6:} This line does not increase the running time as clearly, $\sum_{i=1}^k T(n_i) \leq T(n)$ when $n=\sum_{i=1}^k n_i$.

\noindent {\bf Lines 7--:} It is clear that the worst clause length will decide an overall upper time bound for $Q$. Note that if there are variables left in $F$, then there will be at least two clauses left and one of the cases below must be applicable.

\begin{enumerate}

\item $|w| \geq 5$. Already a rough analysis suffices here: In the call $Q(F(a_1/true))$ $a_1$ as well as all the other variables in $w$ get a fixed value and hence $|w|$ variables are removed. The next call only removes one variable, namely $a_1$. In every of the other $|w|-1$ calls $|w|-1$ variables are removed.
Hence, the running time will be in $O(\tau(|w|,1,(|w|-1)^{|w|-1})^n)$ and the worst case is $O(\tau(5,1,4^4)^n) \subseteq O(1.7921^n)$.

\item $|w|=4$ For a better readability, assume $w=(a \lor b \lor c \lor d)$. As $a$ and $b$ are not singletons there are clauses $a \in y$ and $b \in z$. There are several possibilities for $y$ and $z$, but due to the balanced branching effect, we may disregard cases where $a \in w$ but $\bar a \in y$ etc. 

\begin{enumerate}
\item $y=(a \lor e \lor f \lor g)$, $z=(b \lor h \lor i \lor j)$. The call $Q(F(a/true))$ removes 7 variables -- all variables of $w$ and $y$. The call $Q(F(a/false;b/true))$ removes 7 variables -- all variables of $w$ and $z$. The call $Q(F(a/false;b/false))$ removes 3 variables, because the clause $w=(c \lor d)$ will in the next recursive step be simplified.  The call $Q(F(a/false;b/\bar c))$ removes 3 variables, because the clause $w=(c \lor \bar c \lor d)$ implies $d=$ false, which will be effectuated by the substitution operation.  The call $Q(F(a/false;b/\bar d))$ removes 3 variables for the same reasons. The call $Q(F(a/\bar b))$ removes 3 variables -- $c$ and $d$ must be false. Similarly for the remaining two calls. Hence, the running time is in $O(\tau(7^2,3^6)^n) \subseteq O(1.8348^n)$.

{\em If $|z|=3$, then regardless of $y$ we get cases better than the above case:}

\item $z=(b \lor e \lor f)$. Counting removed variables as previously we get that this case runs in time $O(\tau(6,4^4,3^3)^n) \subseteq O(1.7605^n).$

\item $z=(a \lor b \lor e)$ or $z=(b \lor c \lor e)$ or $z=(b \lor d \lor e)$. All these cases run in time $O(\tau(5^2,4^6)^n) \subseteq O(1.6393^n)$.

{\em If $|y|=3$, then regardless of $y$ we get cases better than the so far worst:}

\item $y=(a \lor e \lor f)$. This case runs in time $O(\tau(6,5,4^3,3^3)^n) \subseteq O(1.7888^n)$.

\item $y=(a \lor b \lor e)$. Already examined.

\item $y=(a \lor c \lor e)$ or $y=(a \lor d \lor e)$.  These cases run in time $O(\tau(5^2,4^5,3)^n) \subseteq O(1.6749^n)$.

{\em If $y$ shares more than one variable with $w$, then regardless of $z$ we get cases better than the so far worst:}

\item $y=(a \lor b \lor c \lor e)$ or $y=(a \lor c \lor d \lor e)$ or $y=(a \lor b \lor e \lor f)$. These cases run in time $O(\tau(5^2,4^6)) \subseteq O(1.6393^n)$, $O(\tau(5^4,4^4)) \subseteq O(1.5971^n)$ and $O(\tau(6^2,5,4,3^4)) \subseteq O(1.7416^n)$, respectively.

\item $y=(a \lor c \lor e \lor f)$ or $y=(a \lor d \lor e \lor f)$. These cases run in time $O(\tau(6,5^2,4,3^4))^n) \subseteq O(1.7549^n)$.

{\em If $z$ shares more than one variable with $w$, then regardless of $y$ we get cases better than the so far worst:}

\item $z=(a \lor b \ldots)$. Already examined.

\item $z=(b \lor c \lor d \lor e)$. This case runs in $O(\tau(5^2,4^6)) \subseteq O(1.6393^n)$.

\item $z=(b \lor c \lor e \lor f)$ or $z=(b \lor d \lor e \lor f)$. These cases run in time $O(\tau(6^2,5,4,3^4))^n) \subseteq O(1.7416^n)$.

\end{enumerate}

\item $|w|=3$. We know that there is another clause $y$ such that $|y|=3$ and $a\in y$ and $y \neq w$. Hence we have a running time in $O(\tau(4,3,2^2)) \subseteq O(1.7107^n)$.

\end{enumerate}
\end{proof}

\section{Conclusions}
We have presented two non-trivial, exact, poly-space algorithms for \hamxsat and provided interesting upper bounds on their running time. Both algorithms point out new interesting research directions and indicate that problems such as \hamsat might be solvable in time better than $O(4^n)$.  Using $P$ as a template when constructing an algorithm for a max Hamming problem, the goal is to analyze the instance at hand to see which calls to the external solver that are superfluous. $Q$ indicates that it is possible to take direct advantage of the inherent structure of the problem itself.

\end{document}